# The Many-Worlds Interpretation of Quantum Mechanics Is Fatally Flawed

Casey Blood
Professor emeritus of physics, Rutgers University
Sarasota, FL
CaseyBlood@gmail.com

## Abstract

The linear mathematics of quantum mechanics gives many versions of reality instead of the single version we perceive, with the version we perceive chosen at random according to a probability law. Because of these peculiarities, the theory requires an interpretation to be fully understood. Over 50 years ago, Everett proposed in his many-worlds interpretation that these characteristics could be accounted for if the mathematics itself, with no collapse or hidden variables, was carefully analyzed. We show this is incorrect because the probability law cannot hold for a system governed by linear equations. Thus the many-worlds interpretation is not viable. Some mechanism, such as collapse or hidden variables, must be added to obtain a satisfactory understanding of the physical universe.

**1. Introduction**

Quantum mechanics, which consists of a set of linear equations for the state vector, gives an accurate description of a wide range of phenomena. But this theory has one most peculiar property; the state vector, which is essentially equivalent to the more familiar wave function, often contains several simultaneously existing versions of reality. In the Schrödinger's cat experiment, for example, the cat is both alive and dead at the same time. Experientially, we perceive only one version—cat alive or cat dead—but quantum mechanics does not indicate which one. Instead, it gives the *probability* of perceiving a particular version.

Thus the mathematics of quantum mechanics seems to leave us with two unanswered questions: Why do we perceive just one version of reality when there are often several in the state vector? And why, if an experiment is repeated many times, are the frequencies of the versions we perceive governed by a probability law? It is this second question which is of primary interest here.

Schemes which attempt to answer these questions are called interpretations of quantum mechanics. One major interpretation is to suppose there are actual particles (or "hidden" particle-like variables), that the particles ride along on and single out just one version of reality, and that it is the singled-out version which we perceive [1,2] . A second major interpretation is to suppose there is some process, outside the conventional laws of quantum mechanics, which collapses the wave function down to just one version [3-5]. If this process collapses the dead-cat version in the

Schrödinger's cat case for example, we would perceive a live cat. There is currently no convincing evidence for either of these interpretations [6].

A third possibility is Everett's [7] many-worlds interpretation (MWI) which claims that a careful examination shows both questions can be answered using only the mathematics of quantum mechanics itself. All that seems to be needed to fully understand the relation between the mathematics and our perceived reality is to give up the idea that there is a single, unique version of each of us. It is this interpretation, currently one of the major possible ways to understand quantum mechanics, which is examined here.

We find the MWI falls short because the probability law, Born's rule, cannot hold in a linear theory. Thus the MWI does not provide us with an acceptable interpretation of quantum mechanics.

## 2. Definition of the many-worlds interpretation

The MWI as we define it is based on a "pure" version of quantum mechanics which we call QM-A. The following principles describe the content and consequences of the mathematics of this unamended, linear, Hilbert space quantum mechanics.

> **(1)** Existence consists solely of the state vectors which obey the usual linear quantum mechanical equations of motion.
> **(2)** There are no particles or hidden variables.
> **(3)** There is no collapse of the state vector, so all the different versions of reality continue forever.
> **(4)** The only "objects" which perceive are the versions of the observer. There is no perceiving "I" in quantum mechanics separate from those versions.
> **(5)** Suppose we have an atomic state $\sum_{i=1}^{n} a_i |i\rangle$ that goes through a detector, and the detector reading is perceived by an observer. Then the equations of motion imply that the full state, after detection and observation, is
> $$\sum_{i=1}^{n} a_i |i\rangle |D,i\rangle |Obs\ perceives\ state\ i\rangle$$
> with version *i* of the observer having a valid brain wave function corresponding to perception of state *i*.

There are also several clarifying remarks. First, the linear mathematics implies there can be no interaction, no means of communication, between the different versions of reality in the state vector. Thus different versions of the observer are not aware of the other versions or what they perceive. Second, the perceptions of which "I" am currently aware correspond to those of a single version of the observer—but there are many equally valid versions of "I". Third, no assumption regarding probability is made in the definition of QM-A (because probability does not directly occur in the linear equations of motion even though conservation of probability—once probability is *assumed*—is implied by the equations). Finally, because we are attempting to explain our perception of results which conform to the probability law, we are concerned here with the probability of *perception* of a given outcome.

Everett claimed that QM-A alone implied all the properties—the perception of only one version of reality, the perception of eigenvalues of operators, agreement among observers, the perception of the exposure of only one film grain by a spread-out wave function, an explanation of the photoelectric effect and other specific phenomena, the probability law, and so on—that are needed to explain our perception of physical existence. (See reference [6] and [8] for a fuller explanation of how most of these properties follow from the mathematics of quantum mechanics.) But we will show here that QM-A cannot account for the probability law.

### 3. Equal validity for all observer states.
We suppose experiments are done on a quantum system with $n$ states,

$$\sum_{i=1}^{n} a(i)|i\rangle, \quad \sum_{i=1}^{n} |a(i)|^2 = 1 \tag{1}$$

There is a detector, D, which registers one of the states, and an observer who perceives the reading on the detector. We do a single run, and consider the system at time 0, after detection by the detector, but before the observer perceives the results. The state is then

$$|\Psi(0)\rangle = \left(\sum_{i=1}^{n} a(i)|i\rangle|D:i\rangle\right)|Obs\ perceives\ no\ reading\rangle \tag{2}$$

Now we consider the system at time $t$, after the observer looks at the reading on the detector,

$$|\Psi(t)\rangle = \sum_{i=1}^{n} a(i) U(t)\{|i\rangle|D:i\rangle|Obs\ perceives\ no\ reading\rangle\}$$

(3)

where $U(t)$ is the linear time translation operator. Under time translation, the observer splits into $n$ versions so the state is

$$|\Psi(t)\rangle = \sum_{i=1}^{n} a(i)|i\rangle|D:i\rangle|ver.\ i\ of\ the\ obs.\ perceives\ reading\ i\rangle \tag{4}$$

(Note that a "preferred basis" was not used in obtaining Eq. (4); it follows directly from the interactions contained in $U(t)$.) The problem is to relate this to probability of perception. Note first that there is no "I," outside the state vectors of quantum mechanics, that is somehow "assigned" to just one of the versions of the observer; only the versions of the observer perceive (principle (4)). Second, we consider in this section the simplest assumption, which is that each version of the observer is equally valid. In that case, there is a valid version of the observer perceiving *each* outcome $i$ on *every* run; that is, there are many equally valid "I's". But when every outcome is (validly) perceived on every run, there can be no *probability* of perception. Thus, the probability law cannot hold if the perceptions of all versions of the observer are equally valid. (See also the end of sec. 5.)

**Justification of equal validity.** First, there is no obvious reason why $U(t)$ should lead to a brain state which is not valid—that is, to a state which cannot correspond to our conscious perceptions.

Second, suppose we do a series of auxiliary "experiments." We set—by hand, not by experiment—the detector so it reads $i$ and then have the observer look at the result. Then we know from everyday observations that we are always aware of the $i$ reading, so we must have

$$U(t)\{|D:i\rangle | Obs\ perceives\ no\ reading\rangle\} = \\ |D:i\rangle\| ver.\ i\ of\ the\ obs.\ perceives\ and\ is\ aware\ of\ reading\ i\rangle \tag{5}$$

Then putting Eq. (5) into Eq. (3), we get

$$|\Psi(t)\rangle = \sum_{i=1}^{n} a(i) |i\rangle |D:i\rangle \\ |ver.\ i\ of\ the\ obs.\ perceives\ and\ is\ aware\ of\ reading\ i\rangle \tag{6}$$

We therefore see that linearity implies "equal validity" (all versions equally aware). And this returns us to Eq. (4), where the perception of every outcome on every run implies there can be no concept of probability in the linear MWI.

## 4. Unequal validity.
We see that if QM-A is to accommodate probability, then not all versions of the observer can have equal validity. How might this come about? First, in order not to tinker with the quantum mechanical mathematics, we suppose that all versions of the observer have a valid *perception* of the results in the sense that every version of the observer has a valid brain wave function, with a set of firing neurons consistent with the readings on the detectors.

But then we suppose that—depending on the particulars of the photon wave functions which travel from the detector to the observer, and the particulars of the brain wave function—the perceptions of the different versions can be either "valid" or "not valid." Or we could use the terms "aware" and "not aware;" or "conscious" and "not conscious" so that some versions are aware (conscious) and others are not. That is, in an attempt to accommodate the probability law within QM-A, we suppose that not all versions of the observer have equal "validity." This would seem to be the only way out of the no-probability bind presented by the "equal validity" situation in Eq. (4).

To see the consequences of this supposition, we again do a single run and obtain a variation of Eq. (4).

$$|\Psi(t)\rangle = \sum_{i=1}^{n} a(i) U(t) \{|i\rangle | D:i\rangle |Obs\ perceives\ no\ reading\rangle\}$$
$$= \sum_{i=1}^{n} a(i) |i\rangle | D:i\rangle | ver.\ i\ of\ the\ obs.\ perceives\ reading\ i,\ with\ validity\ v\rangle \quad (7)$$

$v$ is a variable which has the value 1 if that version of the observer is "valid" and the value 0 if it is not. A value of 0 means that, for whatever reason, related to the details of the wave functions, our "conscious" perceptions cannot correspond to that version of reality on that particular run. This mixing of "conscious" perception with quantum mechanics may seem like a strange concept. But the MWI failed to produce probabilities when all states were equally valid, so if this interpretation is to succeed, the only option left is to use the valid/not valid idea.

What can the variable $v$ depend on? It can depend on the value of $i$. And it can depend on the details of the photon and observer wave functions. But the state that $U(t)$ operates on in Eq. (7) does not depend on the coefficients $a(i)$, so $v$ *cannot* depend on the coefficients.

### 5. The validity variable cannot lead to the probability law in a linear theory.
We now repeat the *n*-state experiment N times, with N large. The state at the end is

$$|\Psi(t)\rangle_N = \sum_{i_1=1}^{n} \ldots \sum_{i_N=1}^{n} a_1^{m_1} \ldots a_n^{m_n}$$
$$U(t)\{|i_1\rangle \ldots |i_N\rangle | D:[i_1 \ldots i_N]\rangle |Obs\ perceives\ no\ reading\rangle\} =$$
$$\sum_{i_1=1}^{n} \ldots \sum_{i_N=1}^{n} a_1^{m_1} \ldots a_n^{m_n} |i_1\rangle \ldots |i_N\rangle | D:[i_1 \ldots i_N]\rangle \quad (8)$$
$$| ver.\ [i_1 \ldots i_N]\ of\ the\ obs.\ perceives\ reading\ [i_1 \ldots i_N]\ with\ validity\ v\rangle$$

where $m_j$ is the number of $|j\rangle$ states in $|i_1\rangle \ldots |i_N\rangle$.

The average number of $j$ states perceived by the $v=1$ versions of the observer will be

$$\overline{N}_j = \frac{\sum_{i_1=1}^{n} \ldots \sum_{i_N=1}^{n} (\delta_{i_1,j} + \ldots + \delta_{i_N,j}) p[i_1,\ldots i_N] v[i_1,\ldots i_N]}{\sum_{i_1=1}^{n} \ldots \sum_{i_N=1}^{n} p[i_1,\ldots i_N] v[i_1,\ldots i_N]} \quad (9)$$

Here $p[i_1,...i_N]$ is the "probability" that "the" observer is in the state corresponding to $[i_1,...i_N]$. But there is no "the" observer; instead, in the MWI, there is a *version* of the observer in *every* state, so the only values for the "probabilities" that make sense are $p[i_1,...i_N]=1$. Therefore if the probability law is to hold, then we must have

$$\frac{\sum_{i_1=1}^{n}...\sum_{i_N=1}^{n}\left(\delta_{i_1,j}+...+\delta_{i_N,j}\right)v[i_1,...i_N]}{\sum_{i_1=1}^{n}...\sum_{i_N=1}^{n}v[i_1,...i_N]}=N|a_j|^2 \qquad (10)$$

However, $v[i_1...i_N]$ is independent of the coefficients $a_i$ so the left-hand-side of the equation is independent of the coefficients. Thus Eq. (10) cannot hold in general. That is, including the possibility of valid and not-valid states cannot rescue the concept of probability in the linear MWI. (Note: The analysis of Sec. 3 corresponds to $v[i_1...i_N]=1$ for all $[i_1...i_N]$.)

## 6. The work of others
Everett [7], the originator of the MWI, tried to finesse the probability issue in the following way: He supposed the observer perceived only the end results, $m$ + spins and $(N-m)$ − spins. The sums of the magnitudes squared of all those states with $m$ + spins is

$$\frac{N!}{m!(N-m)!}|a_+|^{2m}|a_-|^{2(N-m)} \qquad (11)$$

This has a sharp maximum, as a function of $m$, at $m/N=|a_+|^2$. Everett reasoned that *in the limit* as N approached infinity, all those states with $m/N \neq |a_+|^2$ would essentially disappear (be of no consequence because their norms were essentially 0). And thus the only states perceived would be those which obeyed the probability law ($m/N=|a_+|^2$). But if N is finite, then in order to get the probability law, one must assume states with small but non-zero coefficients are not perceived. And that is an extra assumption which is indefensible within QM-A, so Everett's reasoning is insufficient.

Zurek [9-11] claims to have derived the probability law from basic quantum mechanics. But his starting point is not QM-A because he assumes, *a priori*, that probability is a property of the state vector. (This is stated in his Fact 2 in references [9] and [10]: "Given the measured observable, the state of the system *S* is all that is needed … to predict measurement results, including probabilities of outcomes.") This assumption, however, is contradicted by the reasoning of Secs. 3-5, which shows that there can be no concept of probability and certainly no coefficient-dependent probability law, in pure linear equation, Hilbert space quantum mechanics. Thus, in my opinion, Zurek's scheme does not fall within the bounds of the linear Everett many-worlds interpretation. And he does not specify how the principles of Sec. (2) are to be changed to allow probability in his interpretation.

There have been other attempts to derive probability from within the linear MWI. Deutsch [12] uses decision theory while Wallace and Saunders [13,14] use subjective likelihood. But none of these can get around the arguments showing that principles (1) to (5), including linearity, are incompatible with probability.

## 7. Summary

The many-world interpretation (MWI), first suggested by Everett, is based solely on principles deduced from the linear equation-Hilbert space structure of quantum mechanics. There are no particles or hidden variables and there is no collapse. There are several simultaneously existing, equally valid versions of reality in the state vector, each of which contains a version of the observer. Further

- Each version of the observer perceives its respective outcome on every run of the experiment. Thus, because each version of the observer *always* perceives, surely the perceptions of the *versions* of the observers do not obey the probability law.
- *Only* the versions of the observer perceive.
- There is no non-quantum-mechanical observer associated with just one version of the observer. So you cannot say "I" become version 17 of the observer with such and such a probability because in the MWI, "I" (including my physical body and memories) become *every* version of the observer with 100% probability.

Since each version of the observer has a valid brain wave function, corresponding to some pattern of firing neurons, it seems appropriate to assume each version of the observer is equally "valid." In that case, the bulleted properties imply there is no perceiving entity, no "I" in the MWI whose perceptions could be governed by a probabilistic law. Thus, under the "equal validity" assumption, probability of perception does not exist in the MWI.

The "equal validity" assumption can be defended by using the linearity property. But even if we assume non-equal validities—sometimes the brain wave function can be "consciously aware," sometimes not—for the various versions of the observer, *linearity still implies the probability law cannot hold*. Thus the *linear* many-worlds interpretation proposed by Everett is not an acceptable interpretation. Some modification of it, such as collapse or hidden variables or perhaps even a fundamental difference between inanimate objects and sentient beings, must be added to account for the probability law.